\RequirePackage{lineno}
\documentclass[aps,prl,twocolumn,superscriptaddress,amsmath]{revtex4}  
\usepackage{makeidx}

\usepackage[colorlinks,
hyperindex,
colorlinks,%
citecolor=blue,%
linkcolor=blue,%
urlcolor=blue, %
]{hyperref}
\usepackage{float}
\usepackage[T1]{fontenc}
\usepackage{array}
\usepackage{tabularx}
\usepackage{graphicx}  
\graphicspath{{./}}
\usepackage{subfigure}
\usepackage{epstopdf}
\usepackage{dcolumn}   
\usepackage{bm}        
\usepackage{amssymb}   
\usepackage{amsmath}
\usepackage{booktabs}
\usepackage{multirow}
\usepackage{supertabular}
\usepackage{color}
\usepackage{amsmath}
\usepackage{textcomp}

\usepackage{siunitx}
\makeindex

\begin{document}

\title{Phononic thermal transport along graphene grain boundaries}

\author{Zhen Tong}
\affiliation{Shenzhen JL Computational Science and Applied Research Institute, Shenzhen 518131, China.}

\author{Alessandro Pecchia}
\affiliation{CNR-ISMN, Via Salaria km 29.300, Monterotondo 00017, Rome, Italy}

\author{ChiYung Yam}
\affiliation{Shenzhen JL Computational Science and Applied Research Institute, Shenzhen 518131, China.}
\affiliation{Beijing Computational Science Research Center, Beijing 100193, China}

\author{Traian Dumitric\v{a}}
\email{dtraian@umn.edu}
\affiliation{Department of Mechanical Engineering, University of Minnesota, Minnesota 55455, United States of America}

\author{Thomas Frauenheim}
\email{thomas.frauenheim@bccms.uni-bremen.de}
\affiliation{Beijing Computational Science Research Center, Beijing 100193, China}
\affiliation{Shenzhen JL Computational Science and Applied Research Institute, Shenzhen 518131, China.}
\affiliation{Bremen Center for Computational Materials Science, University of Bremen, Bremen 2835, Germany}


\qquad \\
\linespread{1.2}
\begin{abstract}
We reveal that phononic thermal  transport in graphene is not immune to grain boundaries (GBs) aligned along the direction of the temperature gradient.  Non-equilibrium molecular dynamics simulations uncover a large reductions in the phononic thermal conductivity ($\kappa_p$) along linear ultra-narrow GBs comprising periodically-repeating pentagon-heptagon dislocations. Green's function calculations and spectral energy density analysis indicate that $\kappa_p$ is the  complex manifestation of the periodic strain field, which behaves as a reflective diffraction grating with both diffuse and specular phonon reflections, and represents a  source of anharmonic phonon-phonon scattering. Our findings provide new insights into the integrity of the phononic thermal transport in GB graphene.

\end{abstract}

\maketitle


Next generation of high-performance electronics and sensors require materials with high thermal conductivity able to spread effectively the high density of Joule heat generation along and across various thin films and substrates \cite{pop_2000,MOORE2014163}. Two-dimensional  materials like graphene \cite{novoselov_electric_2004} are very attractive for these applications as they are relatively immune to detrimental size effects on basal-plane thermal conductivity. This is because the highly anisotropic phonon group velocity reduces the impact of scattering by the top and bottom surfaces \cite{shi, ni}.  Nevertheless, thermal transport is significantly impacted \cite{Klemens_1955} by the  defects occurring during synthesis. 
In this respect, the widely-used chemical vapor deposition (CVD) \cite{Suneel} unavoidably produces grain boundaries (GBs).  As   domains nucleate randomly on substrates, their CVD growth and coalescence result in formation of GBs \cite{yu2011control,huang_grains_2011,an_domain_2011,yazyev_polycrystalline_2014}. 

GBs are imagined as periodic arrays of  dislocations \cite{PhysRev.78.275}. In graphene, GBs are strings of   pentagon-heptagon (5-7) edge dislocations \cite{edge,huang_grains_2011, zettl,nemes-incze_electronic_2013} and their organization can gives rise to diverse GB shapes. While in general the thermal gradient can have an arbitrary orientation with respect to the GB line \cite{liu_anomalous_2014, fox_thermal_2019},  only transport {\it across} GBs is perceived to significantly impact $\kappa_p$.  Green's function (GF) calculations \cite{serov_effect_2013, sandonas_first-principle-based_2018} obtained that heat transmission across the GB  can be significantly influenced by the GB structure, size, and shape.  Non-equilibrium molecular dynamics (NEMD)  simulations    \cite{bagri_thermal_2011, cao_kapitza_2012,azizi_kapitza_2017} revealed a discontinuity in the temperature ($T$) profile across  GBs and that higher dislocation densities lead to lower $\kappa_p$.

\begin{figure}
\centerline{\includegraphics[width=5.5cm]{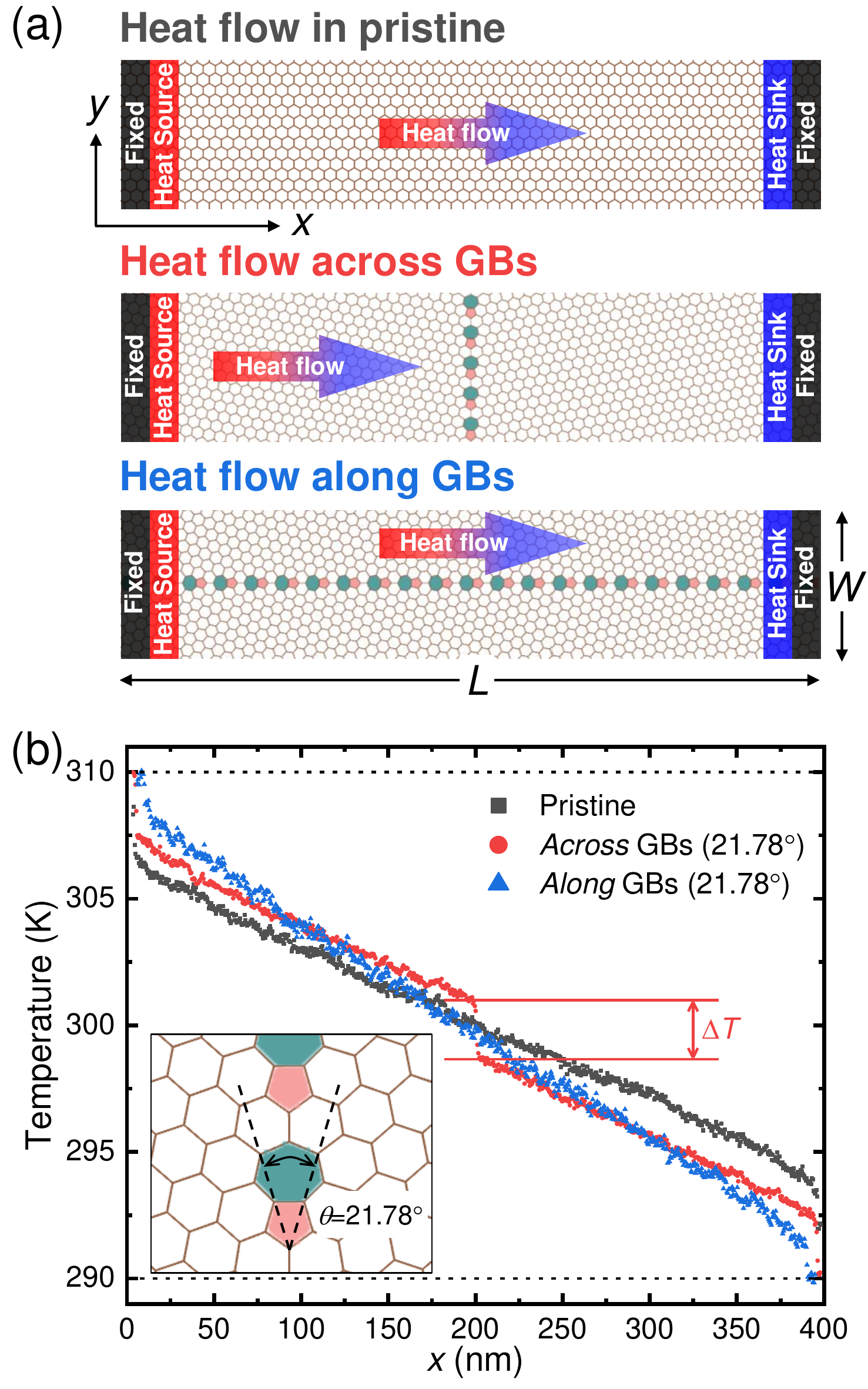}}
\caption{\label{fig:1}
(a) NEMD setup for pristine and GB graphene with $\theta$=21.78$^{o}$.   (b) Computed $T$   profiles.  }
\end{figure} 

\begin{figure*}
	\centerline{\includegraphics[width=12cm]{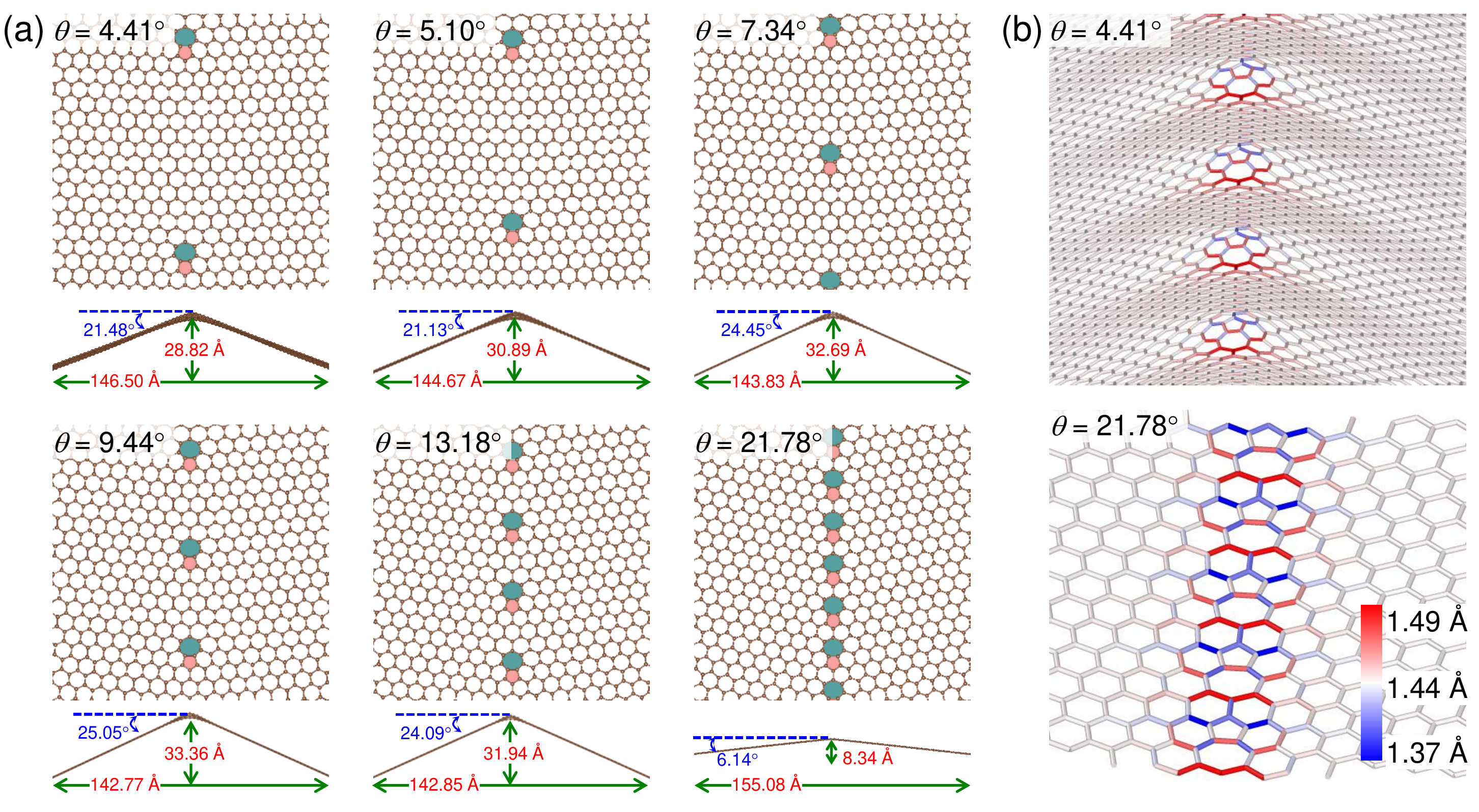}}
	\caption{\label{fig:2}
	(a) The top and side views of 6 considered GBs with different $\theta$. (b) Bird's eye view  along two GB lines.  }
\end{figure*}

In this Letter, we reveal that thermal transport is not immune to GBs oriented {\it along} the thermal gradient. By way of NEMD simulations with LAMMPS \cite{plimpton_fast_1995}, we report $\kappa_p$ reduction along  GBs with  various 5-7 dislocation densities and length scales  $L$ covering ballistic and diffusive transport,  in systems of up to 59,0512 carbon (C) atoms treated with the optimized Tersoff potential \cite{lindsay_optimized_2010}. To gain a clear understanding, we also conducted GF calculations of phononic transmission and conductance, and spectral energy density (SED) calculations to quantify phonon relaxation times.   
The  uncovered $\kappa_p$ behavior with a non-monotonic dependence on 5-7 defect density is unaccounted for by the classical Klemens theory \cite{Klemens_1955}. 

The NEMD setup is presented in Fig. \ref{fig:1}(a). Two-unit cells at each end were kept fixed throughout simulation and ten other neighboring unit cells were designated as {\textquotedblleft}hot{\textquotedblright} and {\textquotedblleft}cold{\textquotedblright} baths maintained at the temperatures $T_h$=310 K  and $T_c$=290 K, respectively.  At steady-state,  the heat flux $\dot{Q}$  was calculated as  the difference of the rate of the kinetic energy extraction from the
two reservoirs $\dot{Q}$ = 0.5<$\dot{Q}_h$-$\dot{Q}_c$>, where $\dot{Q}_h$ and $\dot{Q}_c$ are the instantaneous heat currents flowing into and away from the {\textquotedblleft}hot{\textquotedblright}  and {\textquotedblleft}cold{\textquotedblright} baths. The angular brackets indicate a statistical average taken after the steady state was reached. Graphene edges \cite{barone} can significantly impact thermal transport \cite{liu_anomalous_2014}.  The application of periodic boundary conditions   along $y$ eliminates the lateral edges and allows for the simulation of the thermal transport perpendicular to a single GB line and along antiparallel GB lines (i.e, the 5-7 defect lines run parallel to each other but with opposite directionality)  separated by the lateral periodicity $W$.  Therefore, differences in calculated $T$ profiles,   Fig. \ref{fig:1}(b), can be attributed solely to GBs. 

\begin{figure}
\centering     
\subfigure{\includegraphics[width=5.5cm]{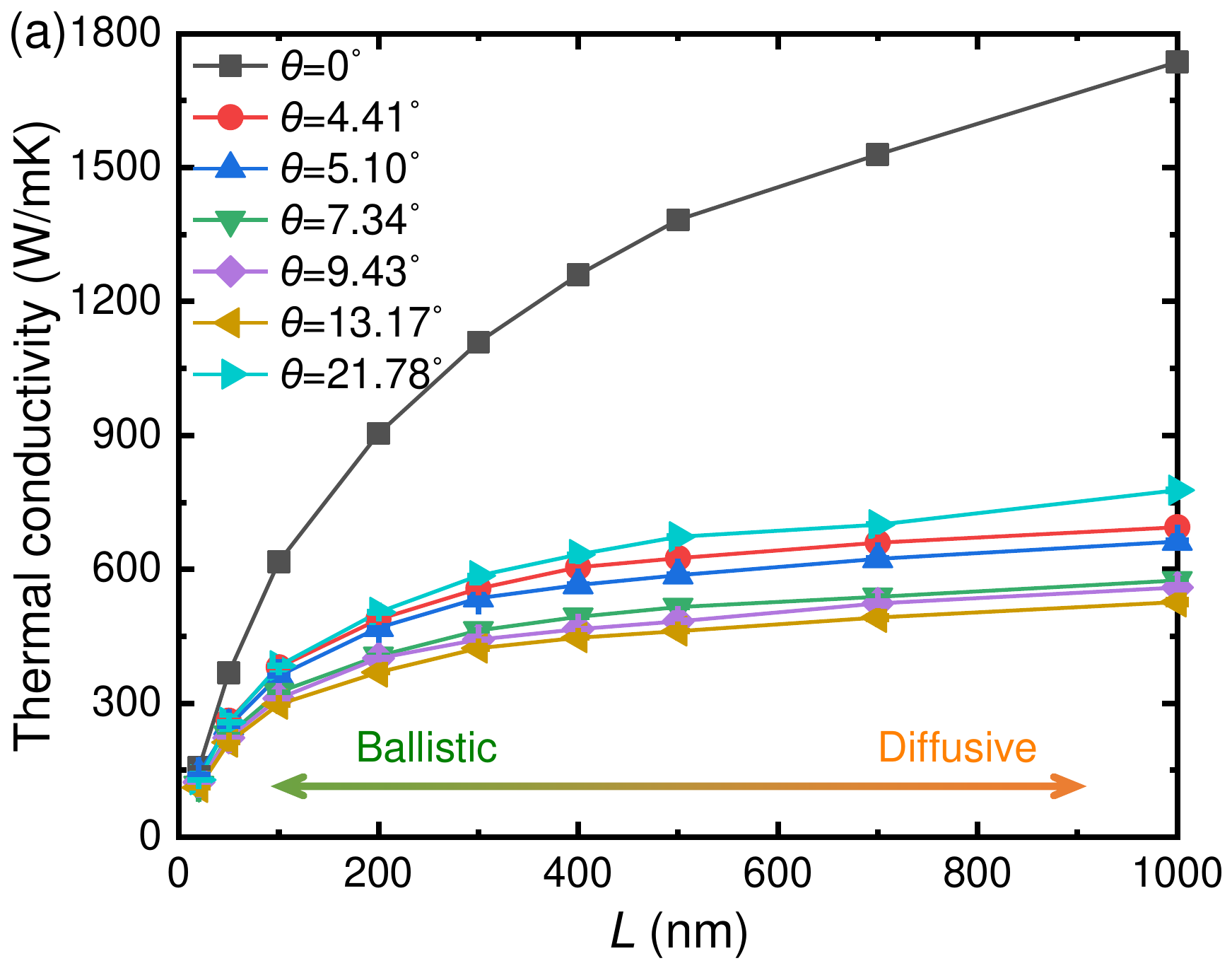}}
\subfigure{\includegraphics[width=5.5cm]{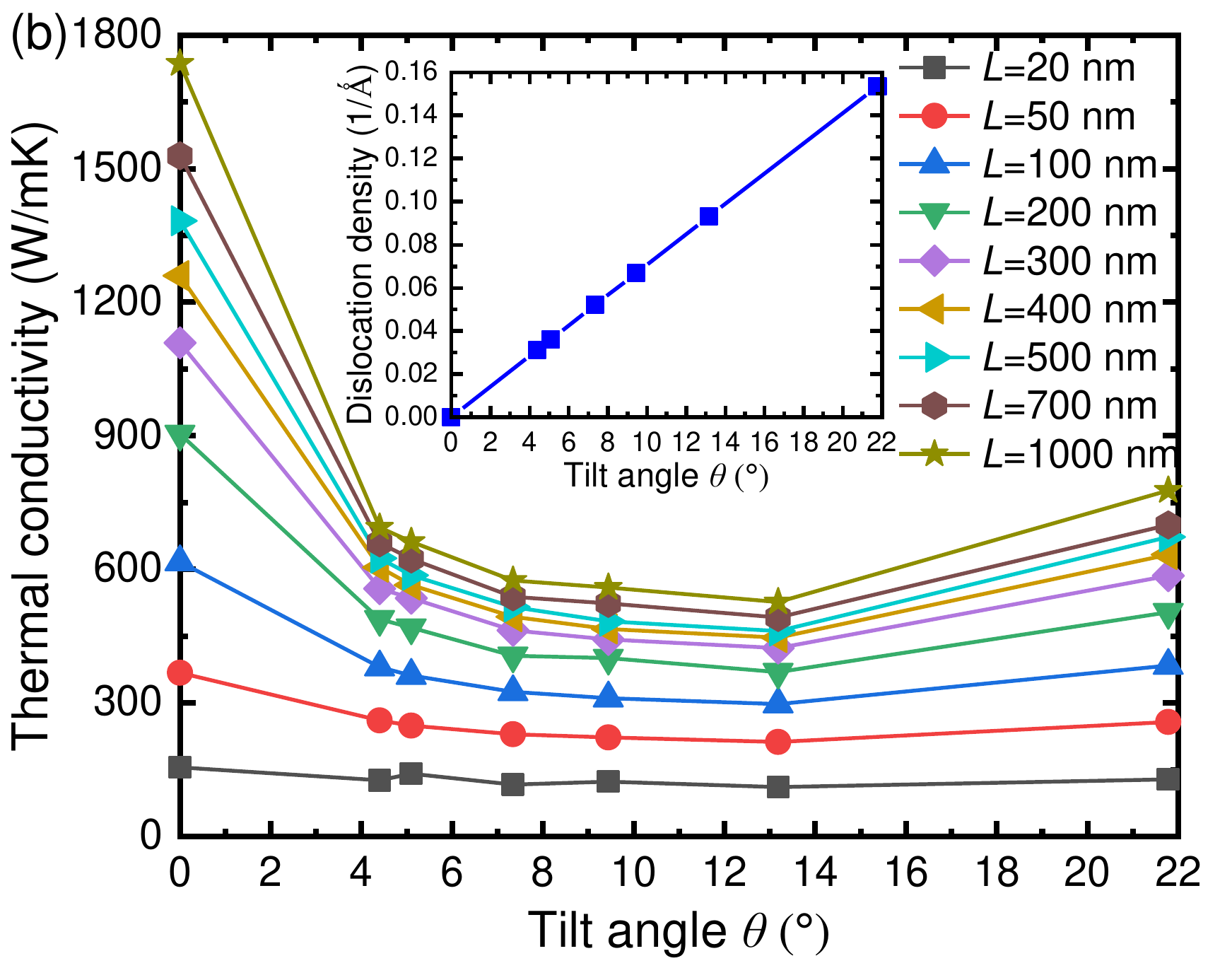}}
\caption{\label{fig:3}
NEMD computed (a)  $\kappa_p$ vs. $L$ at 300 K in graphene and along  GBs with different  $\theta$.   (b) $\kappa_p$ vs. $\theta$ for different $L$. The inset shows the 5-7 density vs. $\theta$. } 
\end{figure}
 
The NEMD calculations of Fig. \ref{fig:1}(b) at $L=400$ nm and $W=15.5$ nm reveal a stark difference in the $T$ profiles across and along the considered GBs, which comprises  aligned 5-7 defects separated by one hexagonal ring. In agreement with Azizi $et$ $al$. \cite{azizi_kapitza_2017}, there is a sharp temperature drop $\Delta T=$2.2 K, corresponding to a thermal resistance $\Delta T/\dot{q}$ of 0.035 Km$^2$/GW across this GB. Here $\dot{q}$ is $\dot{Q}$ per cross-sectional area (defined here based on the 0.33 nm thickness of graphene \cite{mortazavi_thermal_2013,wei_genetic_2020}). Nevertheless, along the GB line, the $T$ profile is smooth and resembles the one obtained for the pristine graphene. By thermal symmetry, the GB lines oriented along the heat flow are adiabatic lines. If heat transfer was purely one-dimensional, then $\kappa_p$ would hardly be impacted along GBs.   Nevertheless, the extracted  $\kappa_p=-\dot{q}(dT/dx)$, reveal a nearly 50\% reduction (633.2 W/mK vs. 1,259.4 W/mK) demonstrating that through the two-dimensionality of the heat transport $\kappa_p$ is significantly impacted even by such linear ultra-narrow GBs.

We have checked the robustness of our result by extensive NEMD, see Fig. S2(b), which considered symmetric tilt GB systems with similar widths $W$ but different $L$  and spread out linear arrangements of the 5-7 defects,  which decrease the tilt angles $\theta$ \cite{yazyev_topological_2010} formed by the crystallographic directions of the neighboring domains,  Fig. \ref{fig:2}(a).  Additionally,  a 5-7 pair introduces local off-plane elevations  \cite{banhart_structural_2011} as a way of reliving the strain stored in the dislocation core.   
The resulting {\textquotedblleft}bumpy{\textquotedblright} landscape with a rather blazed profile is visible in Fig. \ref{fig:2}(b) for the $\theta$=4.41\textdegree{} GB. On the same figure, it can be also seen that the the C-C bond extension and compression deformations are strongly localized around the 5-7 cores. The off-plane displacement are opposite in neighboring GB lines, such stable ripple structures are formed. As shown in the side views of Fig. \ref{fig:2}(a), the ripples acquire significant amplitudes of 15$\pm$1.5\AA. Only tor $\theta$=21.78\textdegree{}, the closeness of the 5-7 cores inhibits their off-plane displacements reducing the ripple amplitude to only 4.17 \AA.
For this case, the C-C bond deformations are continuous along the GB line, Fig. \ref{fig:2}(b).   Overall, in all of the rippled structures of Fig. \ref{fig:2} found by energy minimizations, the C-C bonds away from the GB lines are undeformed; the axial prestrain is also very small; it varies monotonically with $\theta$, from -0.2 \% ($\theta$=4.41\textdegree{}) to 0.1 \% ($\theta$=21.78\textdegree{}), see Fig. S1(b).

Figure \ref{fig:3}(a) demonstrates that the differences  between $\kappa_p$  in pristine ($\theta=0$) \cite{ghosh_extremely_2008, xu_length-dependent_2014} and along GBs with different $\theta$ remain significant at different $L$.  In the pristine case, the initial linear increase of $\kappa_p$  ($L$<100 nm) is a signature of pure ballistic behavior, while the subsequent lowering of the rate of increase in $\kappa_p$ at $L\sim$100 nm signals that the thermal transport enters into a diffusive regime.  $\kappa_p$  is expected to increase in a  logarithmically divergent manner at $L$ much larger than the average phonon mean free path ($\sim$775 nm at $T=$ 300 K) \cite{ghosh_extremely_2008, xu_length-dependent_2014}.  Along GBs, $\kappa_p$ initially increases with a smaller slope than in the pristine case. At larger $L$,  $\kappa_p$ becomes  convergent at $L\sim$400 nm when the diffusive regime settles in. At the largest considered $L=$1 $\mu$m, the $\kappa_p$ reduction is of $\sim$60\% or larger with respect to the pristine case. While the dislocation density increases with $\theta$, $\kappa_p$ displays non-monotonic variations, which are  more pronounced for $L>100$ nm, Fig. \ref{fig:3}(b):  $\kappa_p$ decreases from $\theta=$4.41\textdegree{} up to $\theta$=13.18\textdegree{} but presents an anomalous enhancement at $\theta$=21.78\textdegree{}, where the 5-7 density is largest.
 
To gain insight into the mechanism of $\kappa_p$ reduction, we have  pursued complementary GF investigations. As in NEMD, we have partitioned the system into {\textquotedblleft}hot{\textquotedblright} bath, device region, and {\textquotedblleft}cold{\textquotedblright} bath, Fig. \ref{fig:1}(a), such as  the GB lines extend in all regions. The dynamical matrix $D$ was computed using the same optimized Tersoff potential \cite{lindsay_optimized_2010} with a finite difference scheme of the atomic forces, not accounting for finite temperature phonon softening effects owing to anharmonicity.  We computed the ballistic transport for $\theta$=4.41\textdegree{}, 13.18\textdegree{} and 21.78\textdegree{} and compared them to the pristine graphene. The conductance $g$ is evaluated within the Landauer approach in terms of the  transmission coefficient, $t_{p}(\omega)$, \cite{MedranoSandonas2016} 
\begin{equation}
\label{eq_th}
g=\frac{\hbar^2}{2 \pi k_B T^2} \int{\omega^2 \frac{e^\frac{\hbar \omega}{k_B T}}{(e^\frac{\hbar \omega}{k_B T}-1)^2} t_{p}(\omega) d\omega},
\end{equation}
where $k_B$  and $\hbar=h/2 \pi$ are the Boltzmann and the Planck constants, respectively.  $t_{p}(\omega)$, in turn, is computed  \cite{Sandonas2019} based on $D$, as $t_{p}=Tr[G^r\Gamma_L G^a \Gamma_R]$. The retarded GF is given by $G^r(\omega)=[\omega^2 - D - \Sigma^r_L - \Sigma^r_R]^{-1}$ and $\Gamma_{L,R}$ are the broadening functions, $\Gamma_{L/R}=i[\Sigma_{L/R}^r - \Sigma_{L/R}^a]$, for the  {\textquotedblleft}hot{\textquotedblright} and  {\textquotedblleft}cold{\textquotedblright} contacts.  

\begin{figure}
\centering     
\subfigure{\includegraphics[width=5.5cm]{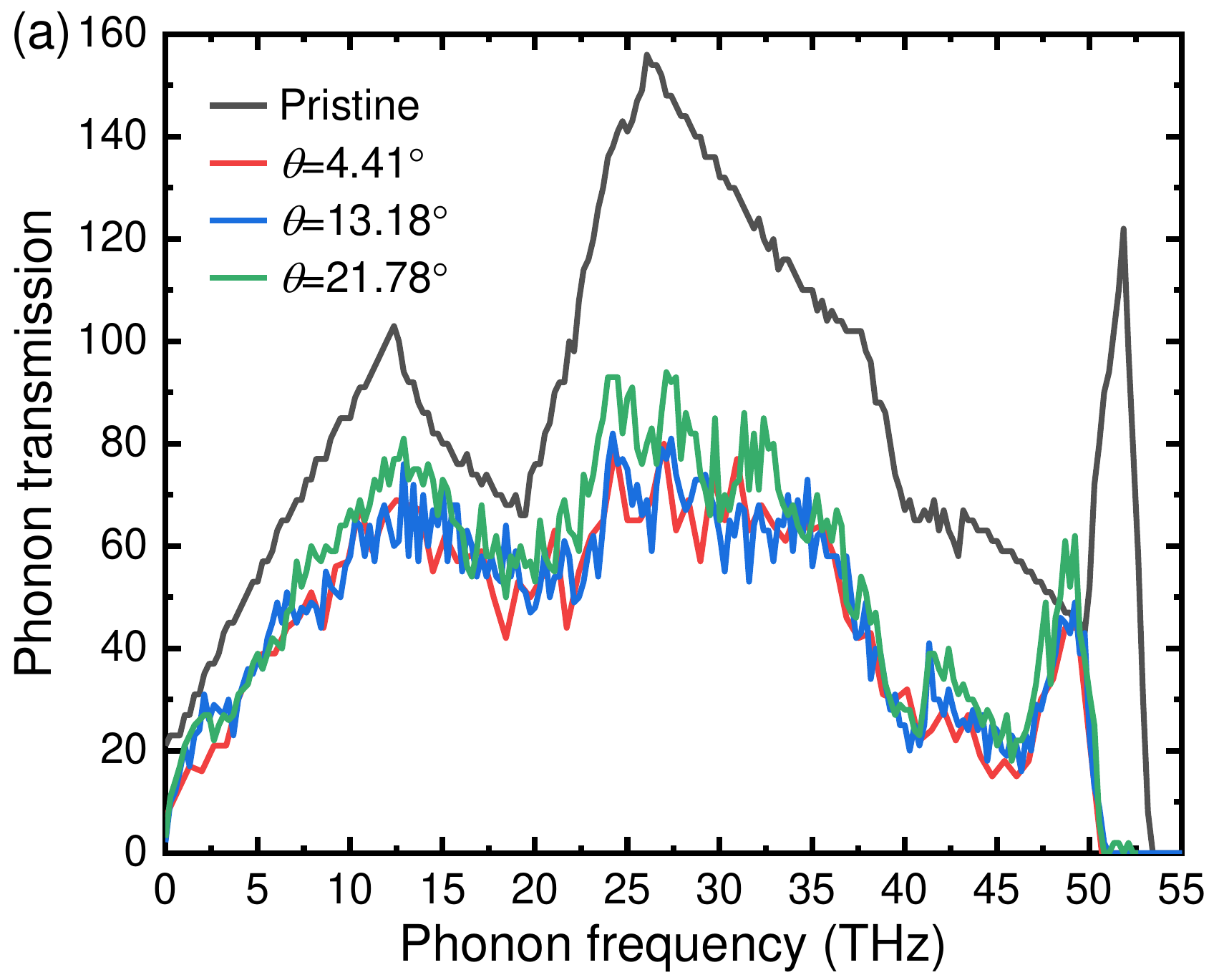}}
\subfigure{\includegraphics[width=5.5cm]{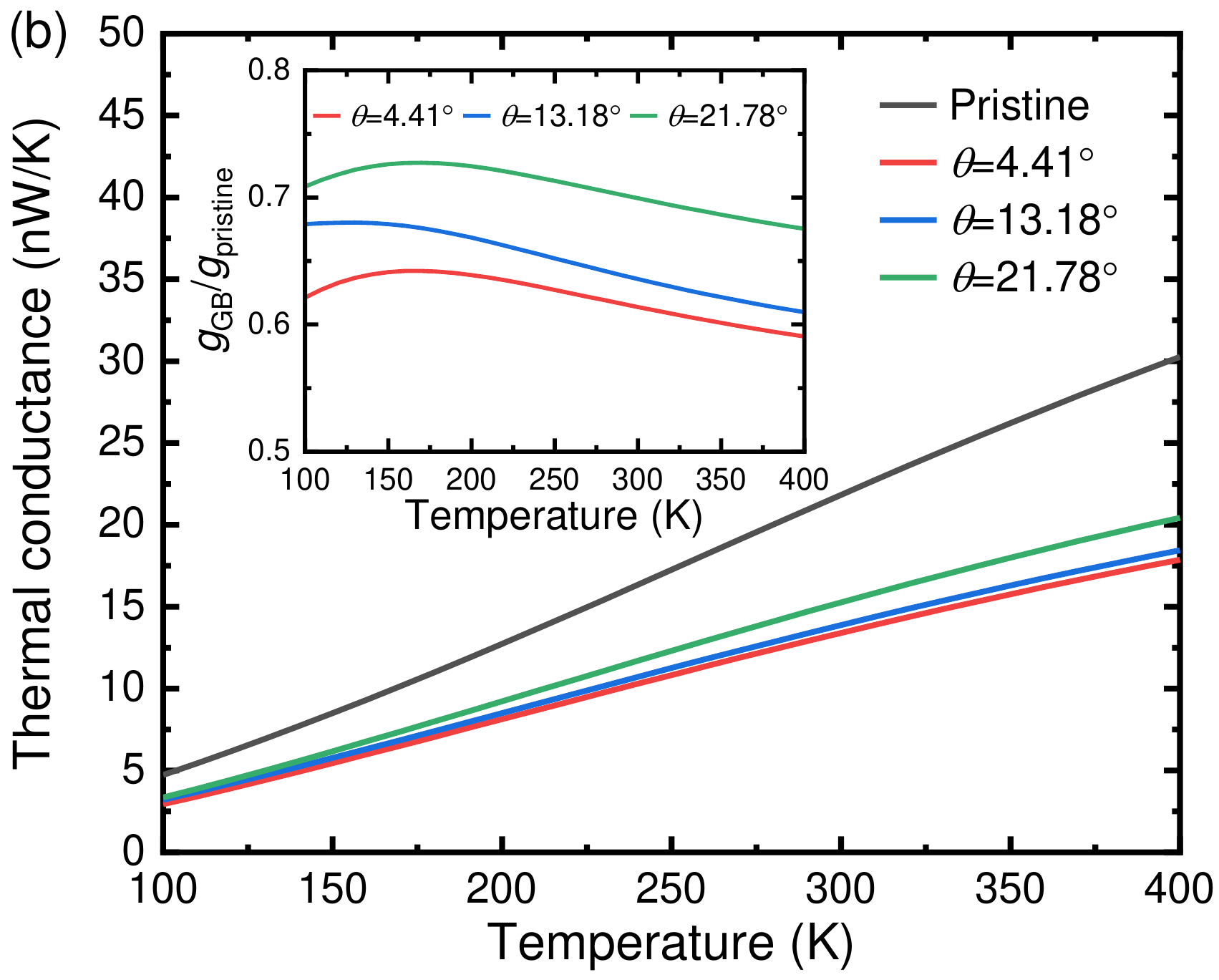}}
\caption{\label{fig:4}
GF computed (a) phonon transmissions  and (b) thermal conductances in pristine  and along GB graphene. Inset shows GB conductances  with respect to graphene.}
\end{figure}

At $L=10$ nm considered here, transport is  coherent and influenced by the elastic scattering onto the GBs. Plots of  $t_{p}(\omega)$ at different  $\theta$ are shown in Fig. \ref{fig:4}(a). Consistent with NEMD,  $t_{p}$ in pristine graphene is higher than in GBs and yields a higher $g$ value, Fig. \ref{fig:4}(b). The conductance is ballistic as reflected in the integer values of $t_{p}$, depending on the number of  phonon states at each $\omega$.  
Surprisingly,  the $t_{p}(\omega)$ reduction by GBs is inversely proportional to the 5-7 density (i.e., $t_{p}$ is largest for $\theta$=21.78\textdegree{}  and smallest for $\theta$=4.41\textdegree{}). 
This dependence uncovers the strain field periodicity effect, which operates as a diffraction grating onto the traveling phonons. Through elastic scattering on the strain around the 5-7 cores, reflective diffraction spectra of various orders take place.  One one hand, for $\theta$=21.78\textdegree{}, strain is continuous along the GB line, Fig. \ref{fig:2}(b). Diffraction is  dominated by the zero-order, which is associated with a specular reflection and larger group velocity. On the other hand, for $\theta$=4.41\textdegree{} GB which has lowest $g$, the 5-7 defects are $\sim$3.2 nm apart.  Destructive interferences introduce stronger phonon localization, which is associated to diffuse reflections and manifests into important thermal resistivity contributions.   
These important higher diffraction orders  are not considered by Klemens \cite{Klemens_1955,Sparavigna}.  This situation reminds of the $\kappa_p$ reduction along central screw dislocations located in nanowires \cite{ni_thermal_2014, xiong}, an effect also not captured by the classical theory \cite{Klemens_1955}.  By introducing periodic nm-scale grooves onto the nanowire surface \cite{Al-Ghalith2016}, $\kappa_p$ could be further reduced through localization of the phonons that were specularely reflected by the dislocation core.
 
In summary, the GF calculations obtained that the resistive contributions caused by the diffuse GB reflection scale inversely with the defect density. The  $g$ reductions are seen also  in the inset of Fig. \ref{fig:4}(b),  which shows $g_{\text{GB}}/g_{\text{pristine}}$  as a function of $T$, with $g_{\text{GB}}$ and $g_{\text{pristine}}$ being $g$ for a given GB and graphene pristine, respectively. 

\begin{figure}
\centering     
\subfigure{\includegraphics[width=5.5cm]{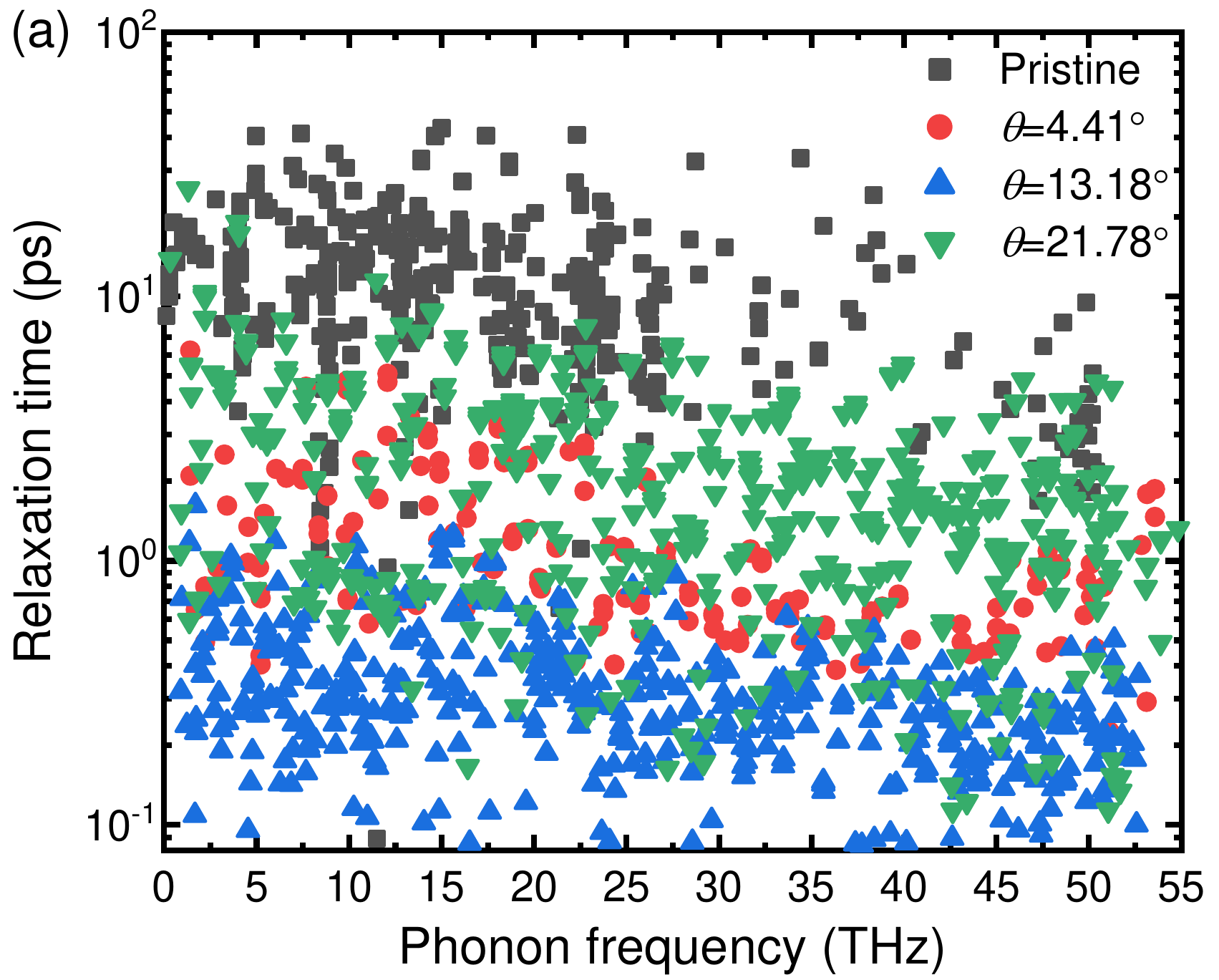}}
\subfigure{\includegraphics[width=5.5cm]{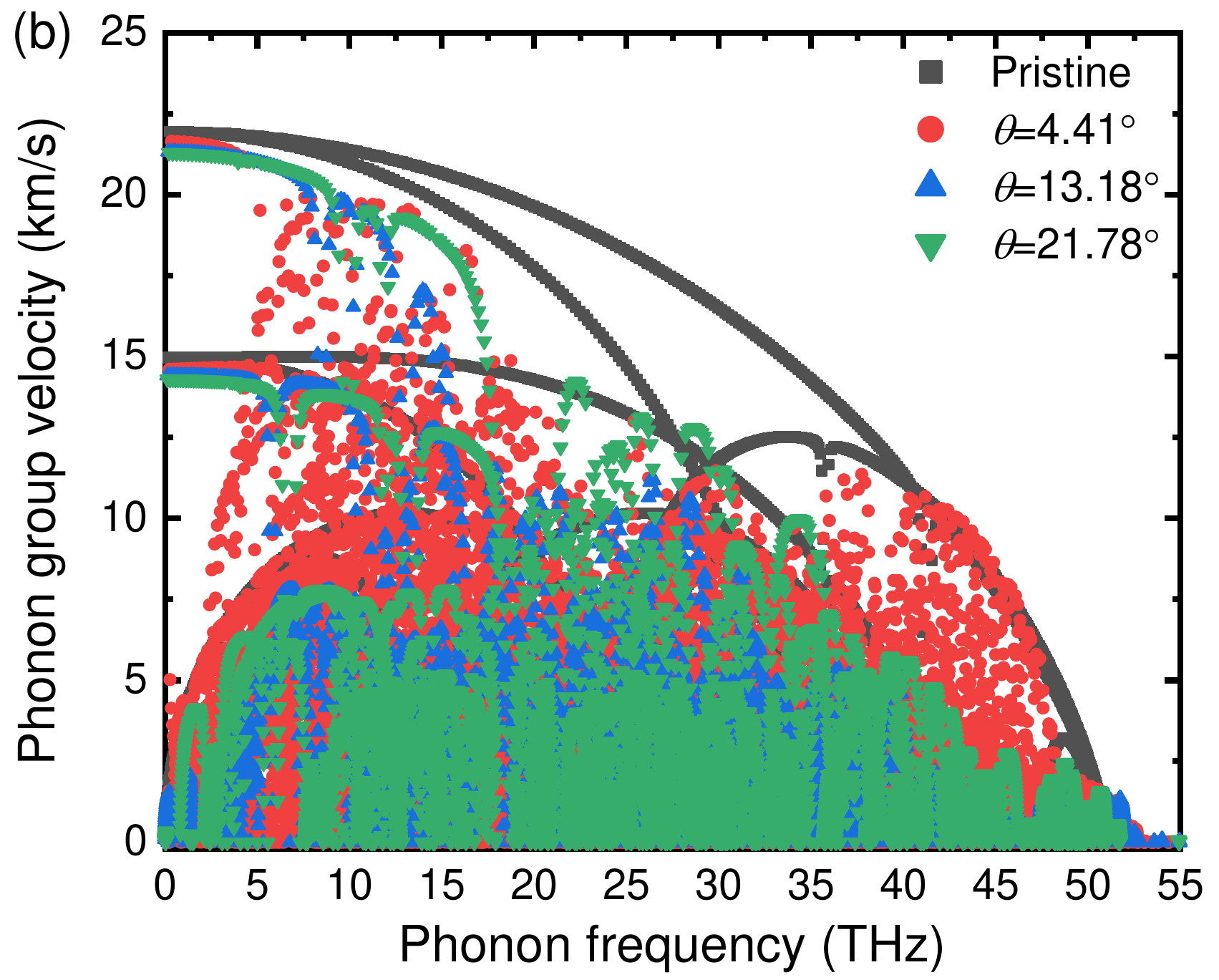}}
\caption{\label{fig:5}
MD calculated (a)  phonon relaxation times and (b) phonon group velocity  in pristine and GB graphene.}
\end{figure}

As transport advances into the diffusive regime, the decay of heat carrying phonons by inelastic scatterings becomes increasingly important.   Recalling that in a phonon gas model, thermal conductivity is  $\kappa_p=\sum_\lambda{c_{\lambda}}v_\lambda ^2\tau_{\lambda}$, where $c_{\lambda}$, $v_{\lambda}$, and $\tau_{\lambda}$ are the specific heat capacity, phonon group velocity, and phonon relaxation time of phonon mode $\lambda$, respectively.    Figure \ref{fig:5}(a) shows $\tau_{\lambda}$, as calculated by SED scheme \cite{thomas_predicting_2010,ni_thermal_2014} and room-temperature equilibrium MD runs. When compared to pristine graphene,  GBs lead to significant  $\tau_{\lambda}$ reductions.  
For the {\textquotedblleft}bumpy{\textquotedblright} GBs,  $\tau_\lambda$  decreases with the increase in defect density. This dependence is opposite to the one for the ballistic phonon transmission delineated above, and explains the crossover in $\kappa_p$  as transport advances into the diffusive regime.  However, for $\theta=$21.78\textdegree{} GB, where  $t_{p}$ is largest compared to other GBs,  we also find the largest $\tau_\lambda$ in Fig. \ref{fig:5}(a).   This concerted behavior explains the consistently larger $\kappa_p$ values for  $\theta=21.78$\textdegree{}  GB with respect to the other considered GBs, Fig. \ref{fig:3}(b). The key role of  $\tau_\lambda$ is further supported in Fig. \ref{fig:5}(b) by the lattice dynamics \cite{gale_general_2003} computed $v_{\lambda}$, which is another key contributor to $\kappa_p$. While for some phonon modes  $v_\lambda$ decreases in $\theta$=21.78\textdegree{} GB, it remains unchanged for the acoustic phonon modes, which are playing a main role in thermal conduction.  Therefore, the weaker phonon scattering in $\theta$=21.78\textdegree{} GB  (as reflected by the larger $\tau_\lambda$) and not an enhancement of $v_{\lambda}$ is the mechanism behind the anomalous $\kappa_p$ behavior.    We associate the weaker anharmonic scattering presented in $\theta$=21.78\textdegree{}  GB to the flatter landscape along the GB line.  The  5-7  off-plane distortions at the other $\theta$ are enhancing anharmonic scattering as they locally couple the in-plane and out-of-plane degrees of freedom \cite{sandonas_first-principle-based_2018}, which are  otherwise  decoupled \cite{bending}.  

In conclusion, we uncovered that through the two-dimensionality of the heat transfer, $\kappa_p$ along linear ultra-narrow GBs is significantly affected. The explanation for the $\kappa_p$ reduction with the defect density goes beyond the simple phonon specular reflection accounted for by Klemens \cite{Klemens_1955,Sparavigna}. $\kappa_p$ alongs GB with 21.78\textdegree{} is largest,  which is opposed to the expected deterioration of thermal transport at larger dislocation density \cite{sandonas_first-principle-based_2018}. The $\kappa_p$ boost is caused by a diffraction grating effect of the GB strain field periodicity, which leads to a specular scattering, and to a reduced anharmonic scattering associated to the flatter GB landscape.  As the 5-7 defects become sparse and {\textquotedblleft}bumpy{\textquotedblright},  elastic scattering on GB becomes diffusive, while anharmonic scattering is enhanced. 
Even for the less dense  case of $\theta=4.14$\textdegree{} GB, where the 5-7 defects are  about 3.2 nm apart, the  $\kappa_p$ reduction remains substantial ( $\sim 60$\% for $L$=1 $\mu$m).  
Our findings provide insights into the thermal response of GB graphene that  are intrinsic to the CVD method \cite{Suneel}, which currently is the main approach to manufacture graphene for high-performance electronics and sensors applications.

\begin{center}
\textbf{SUPPLEMENTARY MATERIAL}
\end{center}
Details on information of GBs, NEMD simulations,  SED calculations, phonon transmission. 

\begin{center}
\textbf{ACKNOWLEDGMENTS}
\end{center}

Simulations were preformed at the Tainhe2-JK of Beijing Computational Science Research Center (CSRC). Z.T. acknowledges the support by China Postdoctoral Science Foundation (Grant No. 2020M680127), Guangdong Basic and Applied Basic Research Foundation (Grants No. 2020A1515110838 and No. 2021A1515011688), and Shenzhen Science and Technology Program (Grant No. RCBS20200714114919142). T.D. and T.F. acknowledge support from DFG FR-2833/7. T.F. acknowledges support from the National Natural Science Foundation of China (Grant No. U1930402).

\begin{center}
\textbf{REFERENCES}
\end{center}

\bibliographystyle{apsrev4_updates}
\bibliography{References}


\end{document}